# Nanoscale Carbon Greatly Enhances Mobility of a Highly Viscous Ionic Liquid

Vitaly V. Chaban[1,2*] and Oleg V. Prezhdo[2]

[1]MEMPHYS – Center for Biomembrane Physics, Syddansk Universitet, Odense M., 5230, Kingdom of Denmark

[2]Department of Chemistry, University of Rochester, Rochester, NY 14627, United States

**Abstract**. Ability to encapsulate molecules is one of the outstanding features of nanotubes. The encapsulation alters physical and chemical properties of both nanotubes and guest species. The latter normally form a separate phase, exhibiting drastically different behavior compared to bulk. Ionic liquids (ILs) and apolar carbon nanotubes (CNTs) are disparate objects; nevertheless, their interaction leads to spontaneous CNT filling with ILs. Moreover, ionic diffusion of highly viscous ILs can increase 5-fold inside CNTs, approaching that of molecular liquids, even though the confined IL phase still contains exclusively ions. We exemplify these unusual effects by computer simulation on a highly hydrophilic, electrostatically structured, and immobile 1-ethyl-3-methylimidazolium chloride, $[C_2C_1IM][Cl]$. Self-diffusion constants and energetic properties provide microscopic interpretation of the observed phenomena. Governed by internal energy and entropy rather than external work, the kinetics of CNT filling is characterized in detail. The significant growth of the IL mobility induced by nanoscale carbon promises important advances in electricity storage devices.

[*] E-mail for correspondence: vvchaban@gmail.com; chaban@sdu.dk.

TOC graphic

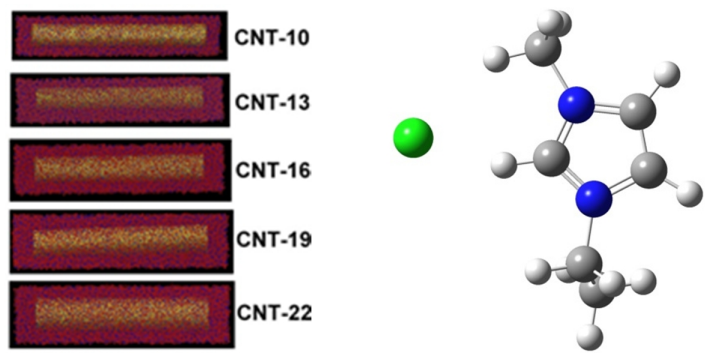

Transport of liquids through carbon nanotubes (CNTs)[1-26] has emerged as an extremely active research area during the last 10-15 years. The goals of nanotechnologists are diverse and well defined. Liquids in nano-channels give rise to the broad field of nanofluidics, including separation membranes, sensors, electrophoretic and thermophoretic channels, gating devices, etc. Nanotubes can be loaded with a solubilized cargo to be delivered inside a living cell. Important features of CNTs include their ability to cross phospholipid membranes and to absorb light that penetrates through living tissues. They give rise to a variety of therapeutic applications and anti-cancer treatments. From the fundamental point of view, understanding liquid behavior in nanoscale spaces is of primary importance, since the resulting properties may be very different from those in bulk liquid, solid and gaseous phases of the same substance. New types of materials can be fabricated inside nanotubes.

Despite a significant amount of reported studies, many pertinent questions remain open. Most reports have been devoted to water, as reviewed thoroughly in Ref.[15] Other confined liquids and gases are represented to a much lesser extent.[27, 28] Hummer and coworkers[20] were among the first to apply molecular dynamics (MD) simulations to characterize a system containing a narrow single-walled armchair CNT. This work introduced a mechanism of pulse-like transmission of water through CNT and visualized empty-filled transitions inside CNT. It was suggested that enhanced water transmission occurs due to tight hydrogen bonding network in the one-dimensional confined water. The rigid CNT carcass shields these hydrogen bonds from fluctuations in the environment. Aquaporin-1, a transmembrane protein, exhibits a comparable water transmission speed.[29]

The theoretical work of Hummer et al.[20, 22] was followed by Majumder and coworkers,[23] who used a membrane consisting of aligned multi-walled CNTs with graphitic inner cores and high area density, crossing a solid polystyrene film. The flows of several liquids, including water, ethanol, hexane and decane, were measured through this membrane at ambient pressure. It was found that the flow rates were 4-5 orders of magnitude faster than the theoretically expected fluid flow.[23] It was hypothesized that the major reason for the reported high fluid velocities was the frictionless surface of the CNT sidewalls.

A series of recent works have been devoted to the discussion of the drastically different structure and dynamics of water, while inside CNTs.[15] The exceptionally fast transport reported initially was questioned in newer publications. For instance, Thomas and coworkers[17] have suggested that the transport enhancement of water is actually much lower. Furthermore, the enhancement in flow decreases significantly with an increase of the CNT diameter. Naguib and coworkers[21] employed transmission electron microscopy, suggesting that fluid mobility inside CNT is reduced rather than enhanced. Ohba and coworkers[30] observed 3-5 times faster transport of water through narrow one-dimensional channels, as compared to wider carbon channels. The latter observation was correlated with a reduced number of hydrogen bonds in confined water films.

In this paper, we investigate a significantly different fluid, compared to water and other molecular liquids. 1-ethyl-3-methylimidazolium chloride, $[C_2C_1IM][Cl]$, is an ionic liquid (IL).[31] It is viscous, immobile and hydrophilic.[32] $[C_2C_1IM][Cl]$ is composed of a complex, asymmetrical cation and a simple anion. Its structure is determined by hydrogen bonds between the chlorine and hydrogen atoms of the imidazole ring of the cation. Rather unexpectedly, our simulations show many similarities between confined $[C_2C_1IM][Cl]$ and water, even though their macroscopic physical chemical properties differ greatly.[32] The interaction of the polar $[C_2C_1IM][Cl]$ and apolar CNTs leads to spontaneous CNT filling. Diffusion of the viscous IL increases 4-5 times, approaching that of molecular liquids, even though the confined IL remains purely ionic. The simulations provide a detailed mechanism of the CNT filling kinetics. The predicted enhancement of the IL mobility carries great promise for batteries and related applications, combining the high surface area of nanoscale carbon with the broad thermal range of IL stability and ionic conductivity. The trends and peculiarities unveiled in the present work using $[C_2C_1IM][Cl]$ as a model, highly viscous IL can be extrapolated to other IL families, especially those containing inorganic anions, such as halogenides, tetrafluoroborates, hexafluorophosphates, etc.

**Results and Discussion**

Five 20 nm long double-walled CNTs (Table 1) were immersed in liquid [C$_2$C$_1$IM][Cl]. Spontaneous filling of these CNT was simulated in the constant pressure constant temperature ensemble, as described in the methodology section. Figure 1 exemplifies atomistic configurations of the filled CNTs.

**Table 1**. Composition of the simulated systems

| System | CNT | CNT diameter, nm | # of confined IL atoms | Density*, atoms×nm$^{-3}$ |
|---|---|---|---|---|
| CNT-10 | (10,10)(13,13) | 1.36 | 1 506 | 85±2 |
| CNT-13 | (13,13)(18,18) | 1.76 | 2 722 | 81±2 |
| CNT-16 | (16,16)(21,21) | 2.17 | 4 522 | 82±2 |
| CNT-19 | (19,19)(24,24) | 2.58 | 6 678 | 82±2 |
| CNT-22 | (22,22)(27,27) | 2.98 | 9 274 | 82±2 |
| BULK | — | — | — | 93.5±0.4 |

\* All densities are given at 363 K (the lowest temperature of spontaneous CNT filling). BULK represents number density of bulk [C$_2$C$_1$IM][Cl].

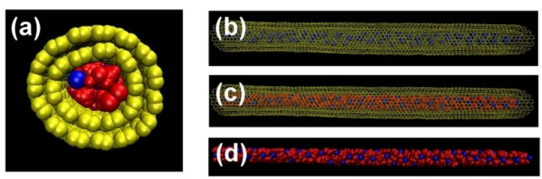

**Figure 1**. Snapshots of the simulated systems containing CNTs. Cations are red balls, anions are blue balls, CNT atoms are in yellow. The snapshots were taken after the filling had been finished. (A) top view of CNT-10; (b) packing of chloride anions in CNT-10, side-view; (c) CNT-13, side view, nanotube atoms are shown as wires to make ions visible; (d) CNT-13, side view, nanotube atoms are omitted for clarity.

The evolution of the energy of interaction between IL and CNT sidewalls during filling (Figures 2 and S2) characterizes the observed non-equilibrium phenomenon at the fundamental level. All curves exhibit three regions: 0 to 6 ns (ions approach the CNT entrance), 5 to 10 ns (ions quickly penetrate the CNT interior), 10 to 15 ns (minor energetic changes due to final re-arrangement of inner and outer ions). Interestingly, before filling, higher curvature of the outer CNT surface gives rise to stronger interaction with the ions (Figure 2). This can appear surprising, since one expects a surface curving away from an ion to interact less, not more. The curvature dependence is inverted after filling, which is also unexpected. A surface curving towards an ion should interact more, not less. The inversion is observed mostly for the cation. The interaction of all tubes with the anion after filling is essentially independent of the curvature. As the relatively bulky $[C_2C_1IM]^+$ particles are adsorbed on the outer CNT surface, a higher curvature introduces smaller perturbation to the cation-anion electrostatic network. Hence, the IL does not need to undergo a major re-organization. The trend inverts when cations are transferred to the confined space, since internal diameters of the smallest CNTs, such as (10,10)(15,15) and (13,13)(18,18), are roughly comparable with the rotational diameter of $[C_2C_1IM]^+$.

CNT filling by the IL increases the CNT-IL binding energies from 1.8-1.9 to 2.6-2.7 kJ mol$^{-1}$ for cations and from 0.23-0.25 to 0.37-0.38 kJ mol$^{-1}$ for anions. Expressed per mole of ions, the energies are higher for the cation (19 atoms) than for the anion (one atom). However, normalized to the total number of adsorbed atoms, the binding energies show that the CNT-chloride interaction is strong. We observe notable differences in IL adsorption in the CNT inner cavity, as compared to polar and apolar molecules, such as methane, ammonia and water. Based on density functional theory calculations with explicit correction for dispersive attraction, these molecules prefer the inner surface of corannulene (μ=2.07 D).[33] The binding energies (enthalpic factor) are 8-16 kJ mol$^{-1}$ higher on the inside than on the outside, due to differences in the strength of interaction between the molecules and the CNT π-electron system. The π-electron interaction plays a less significant role in the case of $[C_2C_1IM][Cl]$, because it is weaker than the ion-ion electrostatic interactions. The ion-ion interactions are responsible for the structure and thermodynamics properties of [$C_2C_1$IM][Cl] both inside and outside CNTs. The main effect of CNTs on the [$C_2C_1$IM][Cl] properties is in creating a repulsive wall that disrupts the ionic interactions.

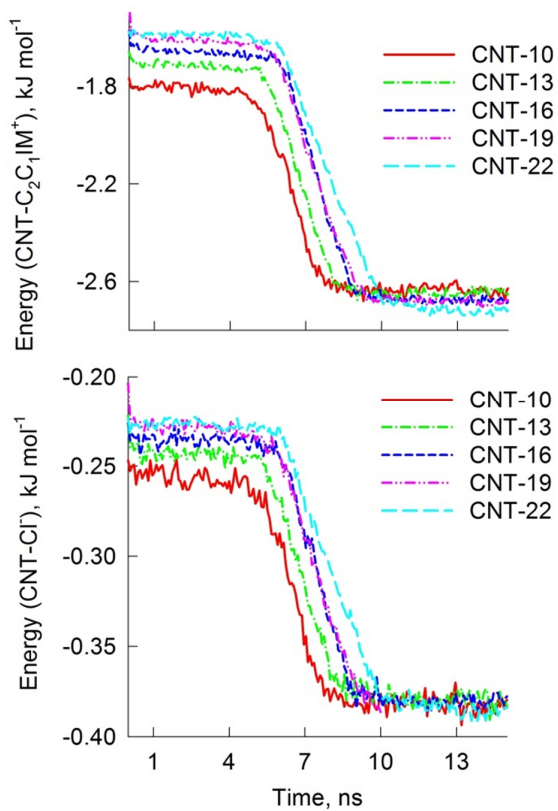

**Figure 2**. Pairwise interaction energy between [$C_2C_1$IM]$^+$ cations and CNTs (top panel); between [Cl]$^-$ anions and CNTs (bottom panel) during CNT filling, conducted at 363 K.

It is not immediately obvious that [$C_2C_1$IM][Cl] should enter pristine CNTs at ambient pressure, since non-modified CNTs are highly hydrophobic and apolar, why ILs are polar and composed of charged particles. The ability of this highly viscous IL to enter even the smallest CNTs is surprising. It has been proven before that CNT can be filled by hydrophilic liquids, such as water, acetonitrile, etc. However, the molecules of these liquids are small, unlike [$C_2C_1$IM][Cl]. They engender low viscosity phases, even near their respective freezing points, whereas [$C_2C_1$IM][Cl] gives rise to a highly viscous, immobile phase at the temperatures of interest. The demonstrated potential of [C$_2$C$_1$IM][Cl] to penetrate into CNTs can likely be used for solubilization of pristine nanotubes. A good performance in solubilizing CNTs has been recently reported for another IL, 1-butyl-3-methylimidazolium tetrafluoroborate.[34] It contains even bulkier ions than [C$_2$C$_1$IM][Cl], but exhibits lower viscosity due to weaker cation-anion electrostatic binding.

Figure 3 investigates the impact of temperature on the filling kinetics, focusing on the smallest CNT (10,10)(15,15) presenting the most interesting case. Importantly, the CNT is not filled at room temperature. Similarly, there is no filling at 320 K under ambient pressure. Additional simulations suggest that filling is prohibited kinetically rather than due to a free energy difference between the initial and final states. Further temperature elevation initiates filling. The filling process is remarkably fast. If CNT is filled at 400 K, it remains filled during 100 ns of simulation with no observable changes in the system enthalpy. The temperature dependence of the filling rate is in satisfactory agreement with the exponential self-diffusion acceleration, predicted by the Arrhenius equation. The filling of CNT (10,10)(13,13) takes ca. 4 ns at 363 K, ca. 2 ns at 400 K and ca. 1.5 ns at 450 K, i.e. with the rates of 0.2, 0.1, and 0.075 ns nm$^{-1}$, respectively. The times were recorded starting from the moment the ions approach the CNT end until stabilization of the CNT-IL interaction energy (Figure 2). The conclusion remains the same for both the cation and anion. The calculated filling times can be easily extrapolated to longer nanotubes if necessary. The short filling time is an important insight of the simulation. The sensitivity of the filling process to temperature indicates that the entropic factor plays a decisive role.

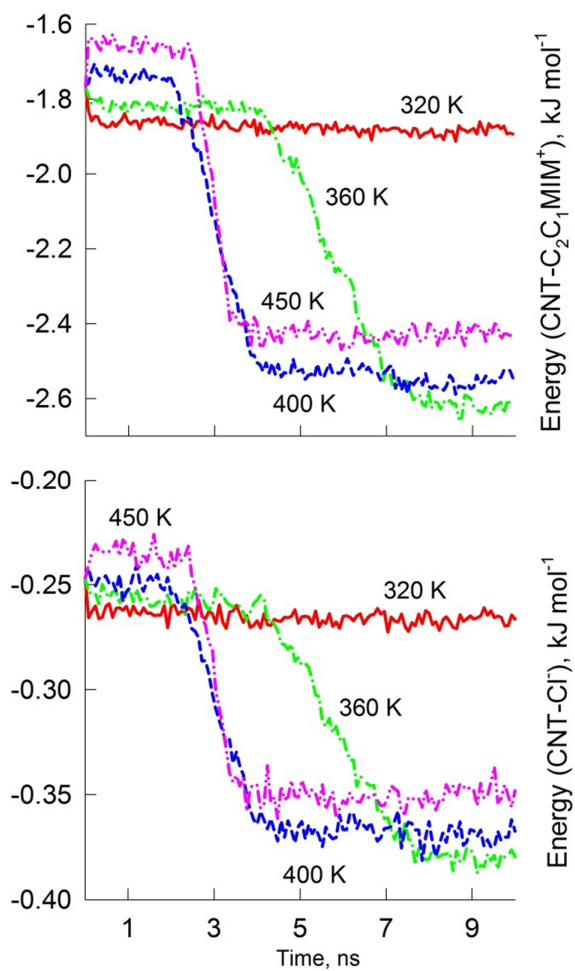

**Figure 3**. Pairwise interaction energy between [C$_2$C$_1$IM]$^+$ cations and CNTs (top panel); between [Cl]$^−$ anions and CNTs (bottom panel) during CNT filling at various temperatures (320, 363, 400, 450 K). Filling of CNT-10 was investigated, as the most interesting case.

The pressure impact on filling (Figure 4) is marginal. The CNT is spontaneously filled at all investigated pressures, including the one that is 100 times smaller than the ambient pressure. The difference between 0.1 and 0.01 bar falls within the simulation uncertainty. The observed trend suggests that all CNTs can be filled even under high vacuum. Insensitivity of the filling process to pressure suggests that filling occurs due to the favorable internal (chemical) energy rather than due to the PdV term. Surface tension provides an alternative descriptor to characterize the thermodynamics of filling for given external conditions.[35]

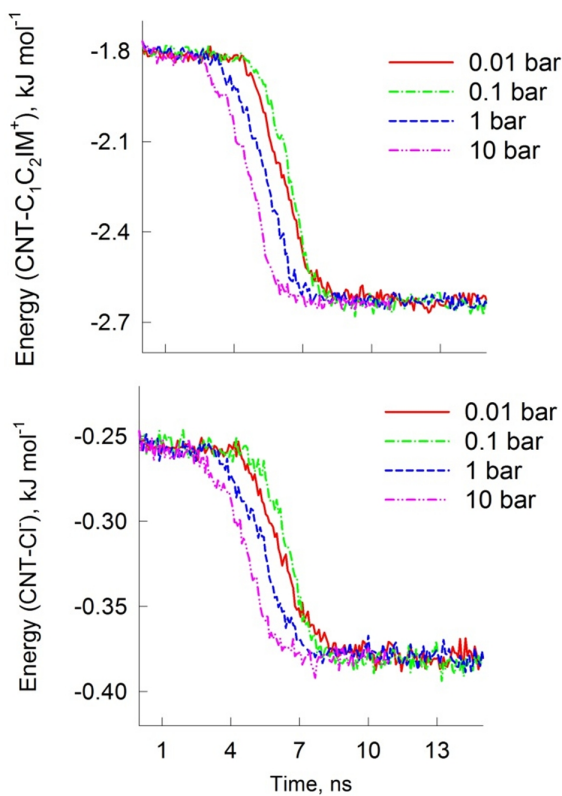

**Figure 4**. Pairwise interaction energy between $[C_2C_1IM]^+$ cations and CNTs (top panel); between $[Cl]^-$ anions and CNTs (bottom panel) during CNT filling at various external pressures (0.01, 0.1, 1, 10 bar). Filling of CNT-10 was investigated at 363 K, as the most interesting case.

Once the filled CNTs had been thermodynamically equilibrated, two subsets of systems were created in the simulation. The first subset represents the single-walled CNTs with the initially selected internal diameters, i.e. CNTs (10,10), (13,13), (16,16), (19,19), and (22,22). The second subset represents the original double-walled CNTs. The ions that did not enter the CNTs during the equilibration (simulated filling) were deleted. The self-diffusion constants of confined $[C_2C_1IM][Cl]$ were computed in all systems (Figure 5). The diffusion constants were determined using the Green-Kubo formula, $D = \frac{1}{3}\int v(0)v(t)dt$, i.e. by integrating the unnormalized linear velocity autocorrelation function. Note that although CNTs are one-dimensional systems formally, ions can diffuse both along and perpendicular to the CNT axis in larger CNTs. Since our goal is to compare IL diffusion inside CNTs to bulk diffusion, we use the expression suitable for bulk. The simulation of self-diffusion was done at 450 K in order to improve sampling and to minimize standard deviations of the computed values.

It may be expected that confined ILs exhibit slower self-diffusion, akin to the recently demonstration for acetonitrile.[27] However, the diffusion constants, D, inside the smallest CNT-10 CNT-13 appear about five times larger than in bulk $[C_2C_1IM][Cl]$, where D=0.2×10$^{-9}$ m$^2$ s$^{-1}$ (450 K). The diffusion constants inside CNT-10 and CNT-13 are nearly the same, whereas D decreases as the CNT diameter increases further. The value inside CNT (22,22) is very close to the bulk value. One can safely assume that further increase of the CNT diameter will produce diffusion constants similar to the bulk value. Diffusion of $[C_2C_1IM][Cl]$ in double-walled CNTs is somewhat slower than in single-walled CNTs. However, the decrease is not drastic. This indicates that the second carbon sidewall plays a minor role in the self-diffusion increase due to confinement. Both single-walled and multi-walled CNTs reproduce the main trend.

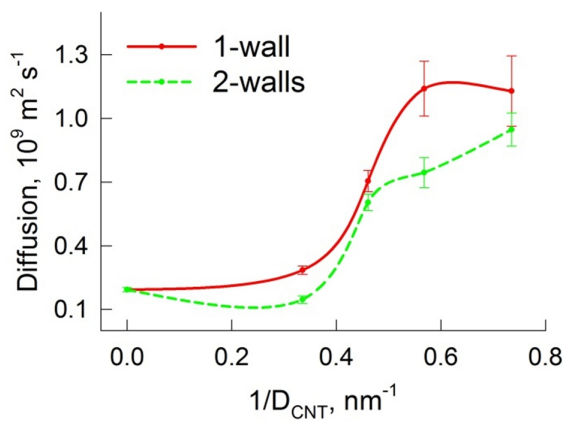

**Figure 5**. Diffusion of $[C_2C_1IM][Cl]$ inside double-walled carbon nanotubes in the CNT-10, CNT-13, CNT-16, CNT-19, CNT-22, and BULK systems. The depicted values are averaged over all cations and anions. Therefore, these values reflect the total mobility of ions in the investigated systems.

In order to provide a microscopic interpretation of the diffusion trend, we perform thermodynamics analysis. Figure 6 depicts the enthalpy of $[C_2C_1IM][Cl]$ confinement and pairwise ion interaction energy. Table 2 decomposes the binding energies into the most important pairwise components. All summarized energies correspond to the temperature of 450 K.

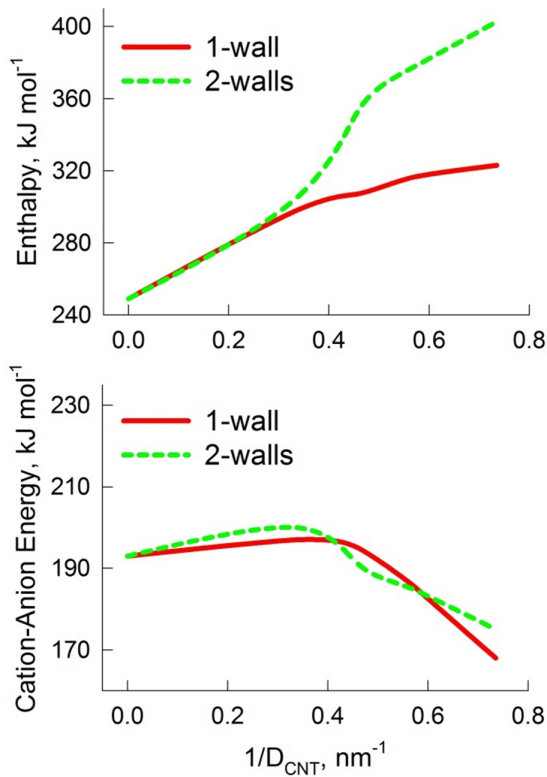

**Figure 6**. (Top panel) Enthalpy corresponding to a single IL ion pair versus CNT diameter and in bulk [$C_2C_1IM$][Cl] ($1/D_{CNT} = 0$). (Bottom panel) Pairwise ion interaction energy versus CNT diameter and in bulk. Both enthalpy and pairwise ion interaction energy are represented as absolute values for clarity.

Due to confinement, electrostatically driven cation-anion attraction decreases as the CNT diameter gets smaller. In the same way, repulsion between the same cations becomes smaller, proportionally. The average CNT-IL binding reaches maximum in small CNTs. Indeed, each ion in these CNTs is located near the CNT sidewall. Although adsorption at the inner carbon sidewalls is driven by the relatively weak van der Waals attraction, the overall adsorption energy is significant (Table 2). In case of the double-walled CNTs, the binding energy is systematically larger. Nevertheless, the CNT-IL interaction is weaker than the cation-anion attraction. All three systems, including smaller CNTs generating drastically accelerated diffusion, exhibit systematically decreased cation-anion interactions.

Importantly, enthalpy in all CNT containing systems exceeds enthalpy in bulk [$C_2C_1$IM][Cl]. Therefore, not only filling of CNTs with [$C_2C_1$IM][Cl] is triggered by the internal energy increase, but also the confined phase of the IL ions inside CNTs is more energetically stable than the corresponding bulk phase. Despite the increased internal energy inside CNTs, ionic self-diffusion in confined conditions exceeds that in the bulk phase. This case is uncommon, since stronger intermolecular interactions usually imply smaller diffusivity. However, thermodynamics does not fully determine transport properties, constituting only an integral measure of chemical forces in a system. The considerations used to explain the water diffusivity increase in small CNTs,[15, 23] apply well to the ionic liquid case. First, chlorine-hydrogen inter-ionic hydrogen bonding is allowed only in one direction out of three in the small CNTs. Second, ideal geometries of single- and double-walled CNTs used in this study favor frictionless movement of ions near the CNT sidewall. All ions in the CNT-10 system and most ions in CNT-13 system (Figure 1) are located near the carbon sidewall. Most studies[10, 22, 36, 37] of water inside CNTs suggest diffusivity increase up to two times, compared to bulk diffusivity. The simulated values depend on the employed force field and, in particular, on the chosen interaction strength between water and $sp^2$ nanotube carbon atoms.[15] In comparison, we report a five-fold acceleration of diffusion in the single-walled CNT-10 and a four-fold acceleration in the double-walled CNT-10. Since the [$C_2C_1$IM][Cl] ion pair is significantly more polar than the water molecule, the diffusivity increase appears dependent on liquid polarity.

**Table 2**. Energetic properties of the studied systems expressed per mole of IL ion pairs

| System | Energy, kJ mol$^{-1}$ | | | |
|---|---|---|---|---|
| | +/− attraction | IL-CNT attraction | +/+ repulsion | −/− repulsion |
| | Single-walled carbon nanotubes | | | |
| CNT-10 | -488 | -91 | +210 | +110 |
| CNT-13 | -613 | -67 | +260 | +165 |
| CNT-16 | -695 | -52 | +295 | +201 |
| CNT-22 | -781 | -35 | +342 | +241 |
| Double-walled carbon nanotubes | | | | |
| CNT-10 | -495 | -160 | +210 | +110 |
| CNT-13 | -605 | -124 | +260 | +160 |
| CNT-16 | -679 | -100 | +295 | +193 |
| CNT-22 | -781 | -38 | +342 | +239 |
| BULK | -943 | — | +470 | +280 |

Can the reported diffusion increase happen due to density differences between the bulk phase of IL and the confined phase? Figure 7 represents relationships between [$C_2C_1$IM][Cl] diffusion, density and external pressure. The latter is responsible for a given density. Little difference in both density and diffusion are observed, when pressure is increased from 0.1 up to 100 bar. Further increase in pressure leads to significant increase in density. Consequently, the density increase correlates with the self-diffusion decrease, by about an order of magnitude at 10 000 bar. An over 10% increase in density due to compressibility is an interesting observation in itself. In the case of low molecular solvents, such as water, a 10% density increase is hard to achieve due to lack of conformational flexibility in these small molecules. Another important conclusion coming from Figure 7 is that [$C_2C_1$IM][Cl] maintains genuine density even at low pressure. Therefore, certain variations of liquid density inside CNT cannot significantly influence the confined ionic self-diffusion.

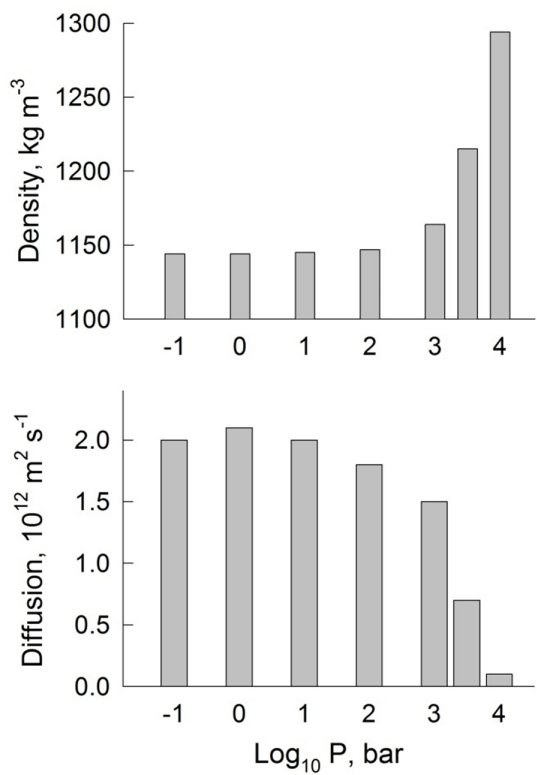

**Figure 7**. (Top panel) density of bulk [$C_2C_1IM$][Cl] vs. external pressure; (bottom panel) diffusion constant versus external pressure. Both density and pressure correspond to MD simulation at 363 K in the constant pressure constant temperature ensemble.

**Conclusions**

We show, for the first time, that a highly viscous IL spontaneously and rapidly penetrates inside CNTs of any diameter at 363 K and higher temperatures. The ability of a highly viscous liquid composed of charged particles to enter even the smallest, apolar CNTs at ambient pressure is surprising. Since the external pressure does not play a significant role in the filling process, filling occurs due to favorable internal energy rather. The sensitivity of the filling process to temperature indicates that the entropic factor plays the decisive role. Despite the increased internal energy inside CNTs, ionic diffusion in confined conditions exceeds that in the bulk phase. This is uncommon, since stronger intermolecular interactions usually imply smaller diffusivity.

The self-diffusivity of [C$_2$C$_1$IM][Cl] inside CNT-10 and CNT-13 is 4-5 times higher than in the bulk phase. In particular, this is in contrast to the decreased diffusivity of acetonitrile and a modest increase for water. The microscopic rationalization of the [C$_2$C$_1$IM][Cl] diffusivity increase is similar to that of water. Since the [C$_2$C$_1$IM][Cl] ion pair is significantly more polar than the water and acetonitrile molecules, the diffusivity increase appears dependent on liquid polarity.

The current study focused on a realistic IL comprising a complex, asymmetrical cation and a simple anion, compared for instance to the recent study employing a simple IL model.[38] It may be interesting to study IL diffusivity in systems with pore size that can accept smaller anions and reject larger cations. The charge separation can result in dramatic changes in diffusion and other properties. The ability of nanoscale carbon to modify significantly the transport behavior of highly viscous ILs can find important applications in chemical technology[39] and nanofluidic devices.

**Methodology**

[C$_2$C$_1$IM][Cl] is selected as a model for viscous ILs. By characterizing the [C$_2$C$_1$IM][Cl]-CNT interactions one can anticipate the corresponding behavior of other ILs with similar transport properties and electronic structure. The thermodynamics and transport properties of [C$_2$C$_1$IM][Cl] IL have been obtained using MD simulations with pairwise potential functions parametrized for [C$_2$C$_1$IM][Cl] by Lopes and Padua.[40] Parameters for non-functionalized CNT ($\sigma$=0.34 nm, $\varepsilon$=0.36 kJ mol$^{-1}$) were taken from the AMBER force field,[41] which is compatible with the IL force field, based on the derivation procedure. Polarizability of ions was included implicitly via the approach developed recently (see Figure S1 and related discussion in Supporting Information).[42, 43] The CNT-IL interactions were simulated using the Lennard-Jones interaction terms with the parameters computed according to the Lorentz-Berthelot combination rules. A comprehensive review of the previous work suggests that this choice provides a reliable and the most widely applied methodology.[15] Polarization of CNT sidewalls by the electric field generated by [$C_2C_1$IM] cations and [Cl] anions is neglected in the present study, since its introduction into the force field is not trivial, requiring additional development and testing. One can expect that CNT polarization should introduce an ion drag, somewhat decreasing the diffusion constants. Polarization effects should be more important in metallic than semiconducting tubes, because the π-electron system of metallic CNTs is more polarizable. Note that short CNTs with metallic (n,m) indices are semiconducting due to finite size effects. Randomly created pores in nanoscale carbon used in applications fall into this category.

20 nm long (10,10)(15,15), (13,13)(18,18), (16,16)(21,21), (19,19)(24,24), and (22,22)(27,27) double-walled CNTs were placed in the orthogonal periodic simulation boxes and surrounded by 1,000-5,000 [$C_2C_1$IM][Cl] ion pairs. Each system was simulated during 10 ns under the corresponding conditions (temperature, pressure). The RTIL-filled configurations (Figure 1) were subject to additional 10×0.5 ns MD runs at 450 K with a time-step of 0.001 ps. The self-diffusion constants were derived applying the Green-Kubo formula to linear velocity autocorrelation function. The ten resulting diffusion constants were processed statistically as results of independent experiments. To ensure such independency, random generation of initial velocities was performed at the beginning of each calculation.

The electrostatic interactions were simulated using direct Coulomb law up to 1.3 nm of separation between interaction sites. The interactions beyond 1.3 nm were computed using the reaction-field-zero (RF0) technique as implemented in the GROMACS simulation suite.[44-46] RF0 provides a description of electrostatic interactions, which is comparable in accuracy to the Particle-Mesh-Ewald approach. RF0 is computationally more preferable is non-isotropic systems. The Lennard-Jones 12-6 interactions were smoothly brought down to zero from 1.1 to 1.2 nm using the classical shifted force technique. The constant temperature was maintained by Bussi-Parrinello velocity rescaling thermostat[47] (with a time constant of 1.0 ps), which provides a correct velocity distribution for a given ensemble. The constant pressure was maintained by

Parrinello-Rahman barostat[48] with time constant of 4.0 ps and compressibility constant of $4.5\times10^{-5}$ bar$^{-1}$. Analysis of structure and thermodynamics was done using supplementary utilities distributed with the GROMACS package[44-46] and in-home tools. All manipulations with nanotubes, such as construction of ideal geometry, division of ions between inner and outer regions, were performed using the set of tools (MDCNT) developed by V.V.C.

All MD simulations were conducted using 12-core nodes of the Exciton cluster (University of Rochester). Domain decomposition technique was used to distribute computational load over the nodes.[45]

**ACKNOWLEDGMENTS**

The reported research was funded in part by grant CHE-1300118 from the US National Science Foundation. MEMPHYS is the Danish National Center of Excellence for Biomembrane Physics. The Center is supported by the Danish National Research Foundation. All computations have been performed on the Exciton cluster hosted at the University of Rochester. We greatly appreciate continuous technical support from the Center for Integrated Research Computing (CIRC).

**SUPPORTING INFORMATION AVAILABLE**

Details of the force field refinement and representation of the data of Figure 2 normalized to CNT surface area can be found in Supporting Information. This information is available free of charge via the Internet at http://pubs.acs.org.

**AUTHOR INFORMATION**

E-mail addresses for correspondence: v.chaban@rochester.edu; vvchaban@gmail.com (V.V.C.)